# A MARVEL-ous Study of How Well Galaxy Shapes Reflect Dark Matter Halo Shapes in Cold Dark Matter Simulations

Blake Keith[1], Ferah Munshi[1], Alyson M. Brooks[2,3], Jordan Van Nest[4], Anna Engelhardt[1], Akaxia Cruz[3], Ben Keller[5], Thomas Quinn[6], and James Wadsley[7]
[1] Department of Physics & Astronomy, George Mason University, 4400 University Drive, Fairfax, VA 22030-4444, USA; bvogel4@gmu.edu
[2] Department of Physics & Astronomy, Rutgers, The State University of New Jersey, 136 Frelinghuysen Road, Piscataway, NJ 08854, USA
[3] Center for Computational Astrophysics, Flatiron Institute, 162 Fifth Ave., New York, NY 10010, USA
[4] Homer L. Dodge Department of Physics & Astronomy, University of Oklahoma, 440 W. Brooks St., Norman, OK 73019, USA
[5] Department of Physics and Materials Science, University of Memphis, 3720 Alumni Avenue, Memphis, TN 38152, USA
[6] Department of Astronomy, University of Washington, Seattle, WA 98195, USA
[7] Department of Physics & Astronomy, McMaster University, ABB-241, 1280 Main Street West, Hamilton, ON L8S 4M1, Canada
Received 2025 January 22; revised 2025 April 17; accepted 2025 April 17; published 2025 June 12

## Abstract

We present a 3D shape analysis of both dark matter (DM) and stellar matter (SM) in simulated dwarf galaxies to determine whether stellar shape traces DM shape. Using 80 central and satellite dwarf galaxies from three simulation suites ("Marvelous Massive Dwarfs," "Marvelous Dwarfs," and the "DC Justice League") spanning stellar masses of $10^6$–$10^{10}\,M_\odot$, we measure 3D shapes through the moment of inertia tensor at twice the effective radius to derive axis ratios ($C/A$ and $B/A$) and triaxiality. We find that stellar shape does follow DM halo shape for our dwarf galaxies. However, the presence of a stellar disk in more massive dwarfs ($M_* \gtrsim 10^{7.5}\,M_\odot$) pulls the distribution of stellar $C/A$ ratios to lower values, while in lower-mass galaxies the gravitational potential remains predominantly shaped by DM. Similarly, stellar triaxiality generally tracks DM triaxiality, with this relationship being particularly strong for nondisky galaxies and weaker in disky systems. These correlations are reinforced by strong alignment between the SM and DM axes, particularly in disk galaxies. Further, we find no detectable difference in either SM or DM shapes when comparing two different supernova feedback implementations, demonstrating that shape measurements are robust to different implementations of baryonic feedback in dwarf galaxies. We also observe that a dwarf galaxy's shape is largely unperturbed by recent mergers. This comprehensive study demonstrates that stellar shape measurements can serve as a reliable tool for inferring DM shapes in dwarf galaxies.

*Unified Astronomy Thesaurus concepts:* Cold dark matter (265); N-body simulations (1083); Hydrodynamical simulations (767); Dwarf galaxies (416); Baryonic dark matter (140); Dark matter (353); Galaxy structure (622); Galaxy dark matter halos (1880)

## 1. Introduction

The dominant mass within the visible radius of dwarf galaxies ($M_* \lesssim 10^{9.5}\,M_\odot$) is in the form of dark matter (DM), making them valuable tests for various DM models. Dwarf galaxies have lower baryon fractions and lower densities compared to their more massive counterparts. This deficiency stems from their shallower gravitational potential wells, as a consequence of their lower halo mass. The weaker gravitational binding makes dwarf galaxies more vulnerable to processes that can remove or prevent the accumulation of baryonic matter, including UV photoionization heating, supernova (SN)-driven outflows, and ram pressure stripping (e.g., L. Mayer et al. 2006; C. Scannapieco et al. 2008; F. Haardt & P. Madau 2012; A. Zolotov et al. 2012; G. A. Dooley et al. 2016; N. Yadav et al. 2017; G. Ogiya 2018; C. M. Simpson et al. 2018; F. Munshi et al. 2021; M. L. M. Collins & J. I. Read 2022). However, we must consider how gravitational coupling between baryons and DM can produce dynamical heating effects that mimic signatures of alternative DM models.

Simulations have revealed that episodic stellar feedback can drive repeated cycles of gas ejection and reaccretion, driving fluctuations in the gravitational potential that gradually transfer energy to DM particles (e.g., J. I. Read & G. Gilmore 2005; A. Pontzen & F. Governato 2012). These fluctuations can transform the central DM density profile from being steeply rising ("cuspy") to being closer to constant ("cored;" J. Peñarrubia et al. 2012; T. K. Chan et al. 2015; R. A. Jackson et al. 2024). This process has been shown to be most effective in dwarf galaxies with stellar masses around $10^9\,M_\odot$ (e.g., A. Di Cintio et al. 2014; A. Lazar et al. 2020; B. Azartash-Namin et al. 2024).

However, self-interacting DM (SIDM) has also been invoked to induce DM density cores in dwarf galaxies (D. N. Spergel & P. J. Steinhardt 2000). Self-interactions can lead to the formation of cores in DM halos through elastic collisions between DM particles, which transfer energy and momentum, effectively heating the central regions and reducing the central density, in contrast to the cusps predicted by cold DM (CDM) DM-only (DMO) simulations (A. Burkert 1995; J. F. Navarro et al. 2010), see also S. Tulin & H.-B. Yu (2018).

Both SIDM and baryonic feedback produce effects that are most pronounced in the central region of galaxies ($r \lesssim 1$ kpc; F. Governato et al. 2010, 2012; A. Pontzen &

F. Governato 2012; V. H. Robles et al. 2017; A. Lazar et al. 2020). This spatial overlap complicates efforts to distinguish between the two models, since both produce cores (F. Governato et al. 2012; A. Pontzen & F. Governato 2012; A. B. Fry et al. 2015; A. A. Dutton et al. 2020). The gravitational interactions between DM and baryons make it particularly challenging to determine whether the CDM or SIDM scenarios better explain the observed stellar distributions in dwarf galaxies (M. Vogelsberger et al. 2014; V. H. Robles et al. 2017; A. Fitts et al. 2019; J. C. Rose et al. 2023). Given these complex interactions, constraining DM models necessitates a dual focus: understanding baryonic processes, and characterizing DM properties.

However, there may be one clear predicted difference between SIDM and CDM with baryonic physics: SIDM DMO simulations produce halos with inner regions that are noticeably more spherical than their CDM DMO counterparts (R. Davé et al. 2001; P. Colín et al. 2002; A. H. G. Peter et al. 2013; V. H. Robles et al. 2017). These shape differences have been used to constrain SIDM in high-mass halos by comparing observed ellipticities at cluster scales to predictions from simulations, as stronger DM self-interactions tend to produce more spherical halo centers (A. H. G. Peter et al. 2013; S. Tulin & H.-B. Yu 2018). While it is easier to probe halo shapes in massive galaxies (via lensing) or in galaxy clusters (via lensing and isophotes), halo shapes have been harder to determine in dwarf galaxies. G. Y. C. Leung et al. (2021) used both gas and stellar kinematics in the dwarf irregular galaxy Wolf–Lundmark–Melotte (WLM) in order to constrain both the central DM density slope and the DM halo shape. They found that WLM seems to be a galaxy with a DM core residing in a prolate halo, hinting at CDM as the underlying model rather than a more spherical halo predicted by SIDM.

In general, only the stellar shape is easy to study in dwarf galaxies. The thinnest galaxies are found at $2 \times 10^9 \ M_\odot$, at which point increasing stellar mass increases the thickness of the disk (R. Sánchez-Janssen et al. 2010). This trend is supported by various studies, including E. Kado-Fong et al. (2020), who used data from the Hyper Suprime-Cam Subaru Strategic Program to show that dwarf galaxies ($10^{7.0} < M_*/M_\odot < 10^{9.6}$) are slightly triaxial (triaxiality $\approx$ 0.75, see Section 3.3), and dwarfs of higher mass exhibit thick disk characteristics at one effective radius but become more spherical at larger radii. S. Roychowdhury et al. (2013) find similar patterns in dwarf irregular galaxies in the Local Volume, where the brightest dwarfs ($B$-band absolute magnitude $M_B \approx -18$) have the thinnest disks. As brightness decreases they have found that dwarf galaxies become increasingly spherical and prolate down to faint galaxies with $M_B = -6.7$. Furthermore, investigations by A. van der Wel et al. (2014) and H. Zhang et al. (2019) have used CANDELS survey data to confirm that low-mass galaxies tend to be systematically rounder compared to their high-mass counterparts.

A complication in using halo shape to discern among DM models is that it is not totally clear how the presence of baryons impacts either model. Baryons and baryonic feedback may also alter the shape of the underlying DM halo. In DMO CDM simulations, inner DM halos ($r \lesssim 1$ kpc or $r \lesssim 10\%$ of the virial radius) consistently appear prolate (J. F. Navarro et al. 1996; B. Moore et al. 2004; J. Bailin & M. Steinmetz 2005; C. A. Vera-Ciro et al. 2014), while the inclusion of baryonic physics leads to rounder inner halo shapes, at least for Milky Way–mass galaxies and above (N. Katz & J. E. Gunn 1991; M. Gustafsson et al. 2006; V. P. Debattista et al. 2008; I. Butsky et al. 2016; V. H. Robles et al. 2017; P. Cataldi et al. 2021; K. T. E. Chua et al. 2022). Notably, K. T. E. Chua et al. (2022) find that dwarf halos remain largely unchanged in shape compared to DMO runs, but a wide variety of baryonic feedback mechanisms all produce rounder shapes in more massive halos compared to DMO results. Additional simulation suites of massive galaxies, including Magneticum, MassiveBlack-II, EAGLE, and cosmo-OWLS, find that DM shapes are consistently more spherical than stellar shapes, having axis ratios ($B/A$ and $C/A$) systematically closer to unity (A. Tenneti et al. 2014; M. Velliscig et al. 2015; L. M. Valenzuela et al. 2024). They also found that the alignment between the stellar and DM shapes becomes more pronounced with increasing mass and for late-type galaxies.

At the other end of the mass scale, studies of the 3D shapes of ultrafaint dwarf galaxies in the EDGE simulation suite have revealed trends with gas content (M. P. Rey et al. 2019; O. Agertz et al. 2020; A. Pontzen et al. 2021; M. D. A. Orkney et al. 2023). Gas-poor ultrafaint dwarfs were found to maintain prolate DM halos, while gas-rich ultrafaints developed rounder, more oblate shapes. This relationship between gas content and halo shape extends to other dwarf galaxy types. For instance, dwarf spheroidal galaxies (dSphs), known to be relatively isotropic and supported by velocity dispersion rather than rotation (M. L. Mateo 1998; M. G. Walker et al. 2009; X.-Z. Er 2013; A. Burkert 2015), are expected to closely follow the underlying DM halo shape with minimal impact from baryons. In a study combining observations and simulations of dSphs within the Local Group (LG), W. L. Xu & L. Randall (2020) inferred galaxy shapes from the correlation of the projected ellipticity and the central surface brightness. They found two populations of dwarfs with different mass-to-light ratios, representing oblate and prolate shapes, which are not fully characterized with their analysis of the FIRE simulations (A. Fitts et al. 2017; P. F. Hopkins et al. 2018) that produced prolate shapes.

We extend these works over a broad range of simulated dwarf galaxy stellar masses, $10^6$–$10^{10} \ M_\odot$. Our goal is primarily to understand in what range of masses and in what type of galaxies the stellar shapes trace the underlying DM halo shape, in order to understand if stellar shapes can be used to infer the underlying DM model. We demonstrate that stellar shape can be used to infer DM shape, particularly in lower-mass galaxies lacking a stellar disk ($M_* \lesssim 10^{7.5} \ M_\odot$). The paper is organized as follows. In Section 2, we describe the simulations used for this work and the methods used to obtain the shapes. In Section 3, we present the axis ratio distributions of DM and stars and compare the shape distributions for galaxies with and without a stellar disk, highlighting the implications for inferring DM shapes. In Section 4, we discuss the influence of baryonic feedback on DM shapes and the implications of using different SN feedback models. Finally, we summarize our conclusions in Section 5.

## 2. Simulations and Methods

We analyze a subset of 80 galaxies, including 15 satellites, selected from the "Marvel-ous Dwarfs" (J. M. Bellovary et al. 2019; F. Munshi et al. 2019; F. Munshi et al. 2021), "DC Justice League" (E. Applebaum et al. 2021; F. Munshi et al.

2021), and "Marvelous Massive Dwarf Zooms" simulation suites hereafter "Marvel," "DCJL," and "Massive Dwarfs," respectively. Each suite consists of cosmological zoom-in simulations, with high-resolution regions embedded in larger, low-resolution volumes. Marvel, DCJL, and Massive Dwarfs were all run with the $N$-body + smoothed particle hydrodynamics code CHANGA (H. Menon et al. 2015), which utilizes the hydrodynamic modules from GASOLINE2 (J. W. Wadsley et al. 2004, 2017).

The Marvel suite focuses on volumes specifically selected to contain dwarf galaxies that are isolated from massive neighbors (∼2–7 Mpc), using a WMAP3 cosmology (D. N. Spergel et al. 2007). These simulations implement gas, initial star, and DM particle masses of 1410 $M_\odot$, 420 $M_\odot$, and 6650 $M_\odot$, respectively, and have a gravitational force resolution of 60 pc. The DCJL suite simulates ∼1 Mpc regions centered on Milky Way–mass galaxies using a Planck cosmology (Planck Collaboration et al. 2016). These slightly lower, "Near Mint" resolution simulations implement initial gas, initial star, and DM particle masses of $2.7 \times 10^4$ $M_\odot$, $8 \times 10^3$ $M_\odot$, and $4.2 \times 10^4$ $M_\odot$, respectively, and have a force resolution of 170 pc. The Massive Dwarfs suite contains zoom-in simulations of isolated dwarf galaxies, with initial conditions drawn from the "Romulus25" (M. Tremmel et al. 2017) simulation volume with initial gas, star, and DM particle masses of $3.3 \times 10^3$ $M_\odot$, 994 $M_\odot$, and $1.8 \times 10^4$ $M_\odot$, respectively, and achieve a force resolution of 87 pc (Mint Resolution).

### 2.1. Star Formation and Gas Cooling

Marvel, DCJL, and Massive Dwarfs utilize a probabilistic $H_2$ abundance star formation scheme and gas cooling (C. Christensen et al. 2012), metal line cooling and metal diffusion (S. Shen et al. 2010), and nonequilibrium formation and destruction of molecular hydrogen. We adopt a spatially uniform, time-dependent cosmological UV background following F. Haardt & P. Madau (2012) in order to model photoionization and heating.

Star formation occurs stochastically in the presence of molecular hydrogen, when the gas is sufficiently cold ($T < 1000$ K) and dense ($n > 100 m_H$ cm$^{-3}$). The probability of a star particle forming within a time $\Delta t$ is given by

$$p = \frac{m_{gas}}{m_{star}}(1 - e^{-c_* X_{H_2} \Delta t / t_{form}}), \quad (1)$$

where $m_{gas}$ is the mass of the gas particle, $m_{star}$ is the initial mass of the forming star particle, $t_{form}$ is the local dynamical time, and $c_* = 0.1$ is the star formation efficiency parameter. $c_*$ combined with the fraction of nonionized hydrogen in $H_2$ ($X_{H_2}$) reproduces the normalization of the Kennicutt–Schmidt relation (C. R. Christensen et al. 2014).

SN feedback in Marvel and DCJL is governed by a "blastwave" model (G. Stinson et al. 2006), where thermal energy, mass, and metals are deposited into neighboring gas following SNe, where only stars with masses between 8 $M_\odot$ and 40 $M_\odot$ are assumed to produce Type II SNe. The energy coupled to the gas per SN is $1.5 \times 10^{51}$ erg. Following a SN, gas cooling is shut off for a period of time equal to the momentum-conserving "snowplow phase" (C. F. McKee & J. P. Ostriker 1977). The combined processes emulate the energy deposited within the interstellar medium by all processes related to young stars, including UV radiation (see J. H. Wise et al. 2012; O. Agertz et al. 2013).

SN feedback in Massive Dwarfs is governed by the "superbubble" model (B. W. Keller et al. 2014), which simulates the combined energy and momentum from clustered SN events. The energy coupled to the gas per SN is $10^{51}$ erg. In contrast to blastwave models, which are resolution dependent and often fail to account for natural processes such as clustered star formation, the superbubble model addresses these issues with subgrid models to describe the merging and thermalization of individual winds and SN energies to form superbubbles (B. B. Nath & Y. Shchekinov 2013; N. Yadav et al. 2017). This model incorporates thermal conduction to produce the expected interior densities of superbubbles by accounting for temperature gradients and evaporation processes. When superbubbles are too small to be fully resolved, a separate treatment for hot and cold gas phases is used to prevent overcooling (B. W. Keller et al. 2014, 2015, 2016).

Marvel, DCJL, and Massive Dwarfs all implement massive black hole (MBH) physics as presented in M. Tremmel et al. (2017; see also J. M. Bellovary et al. 2019, for additional details). We briefly summarize this modeling here. MBHs form in overdense regions (3000 cm$^{-3}$ in Marvel and Massive Dwarfs, and 15,000 cm$^{-3}$ in DCJL), in gas with low metallicity ($Z < 10^{-4}$), $H_2$ fractions below $10^{-4}$, and a maximum temperature of $2 \times 10^4$ K. When these conditions are met, seed MBHs form with masses of approximately 25,000 $M_\odot$ in Marvel, $10^5$ $M_\odot$ in Massive Dwarfs, and 50,000 $M_\odot$ in DCJL. MBHs grow by accreting gas from nearby gas particles, and distribute thermal energy into neighboring gas particles. The feedback energy is tied to the accretion rate, where we assume a radiative efficiency $\varepsilon_r = 0.1$ and a feedback coupling efficiency $\varepsilon_f = 0.02$. These simulations also incorporate a subgrid model for producing dynamical friction as described in M. Tremmel et al. (2015), which is capable of tracking locations of MBHs in their halos and modeling MBH mergers. These combined MBH subgrid models have been shown to be successful at reproducing galaxy scaling relations and properties from observations (M. Tremmel et al. 2017; J. M. Bellovary et al. 2019).

### 2.2. Halo Identification

Amiga Halo Finder (S. R. Knollmann & A. Knebe 2009) was applied to the Marvel, DCJL, and Massive Dwarfs simulations to identify DM halos, subhalos, and the baryon content within. We identify the virial radius as the radius at which the halo density is 200 times the critical density of the Universe at a given redshift.

### 2.3. Intrinsic 3D Shapes

For each component of a galaxy (DM and stellar matter (SM)), a thin ellipsoid shape is calculated through the shape tensor (B. Allgood et al. 2006, or some other form of the moment of inertia tensor) defined by

$$\mathcal{S}_{i,j} = \frac{1}{M} \sum_k m_k r_{k,i} r_{k,j}, \quad (2)$$

where $m_k$ is the mass of the $k$th particle, $r_{k,i}$ and $r_{k,j}$ are, respectively, the $i$ and $j$ components of the position vector of the $k$th particle relative to the center of the galaxy, and $M$ is the total mass in the considered volume. Following the iterative algorithm outlined in M. Tomassetti et al. ([2016](#)), we calculate $S$ in radial shells where its eigenvalues ($\lambda_i$) yield the principle axes of the shell:

$$i = \sqrt{3 * \lambda_i} \text{ for } i = A, B, C \text{ and } \lambda_A \geqslant \lambda_B \geqslant \lambda_C. \quad (3)$$

The principal axes $A$, $B$, and $C$ correspond to the semimajor, semi-intermediate, and semiminor axes of an ellipsoid, respectively. We consider the algorithm to have converged when the change in axis ratios between iterations is less than 0.001 for all axes. This process yields the intrinsic 3D axis ratios $Q = B/A$ and $S = C/A$ for each galaxy as a function of radius. To minimize the impact of local substructure and noise, we fit the axis ratios as a function of radius, $Q(r)$ and $S(r)$, with polynomials with order = 3–5 before taking measurements at both radii. These ratios provide a quantitative measure of the galaxy's shape: $Q = 1$ indicates perfect circular symmetry in the $B$–$A$ plane (likely aligned with the rotation of the galaxy; V. P. Debattista et al. [2008](#); M. D. A. Orkney et al. [2023](#)), while smaller values of $Q$ indicate increasing elongation. Similarly, $S \approx Q \approx 1$ would indicate the galaxy has an overall spherical shape, while small values of $S$ describe increasingly "flattened" galaxies. For example, a thin disk galaxy would have $S \approx 0.2$, while an elliptical galaxy typically has higher values of both $Q$ and $S$. To compare the axis ratios of stellar and DM components, we calculate the ratios $Q_{DM}/Q_*$ and $S_{DM}/S_*$.

Following recommendations from M. Zemp et al. ([2011](#)), we implement limits on the number of particles per bin in order to achieve high quality local ellipsoid fits. M. Zemp et al. ([2011](#)) recommend having $\mathcal{O}(1000)$ particles in each radial bin to ensure the iterative fitting method converges, though C. A. Vera-Ciro et al. ([2014](#)) find convergence accurate to 1% for as few as 200 particles. Furthermore, M. S. Fischer & L. M. Valenzuela ([2023](#)) demonstrate that shape measurements are undefined for regions of near constant density. As our primary measurements are taken at 2 $R_{eff}$, a higher radius than a typical DM core at approximately the size of the galaxy half-light radius (J. I. Read et al. [2016](#); A. Fitts et al. [2017](#); A. Fitts et al. [2019](#)), we expect that our shape measurements are well defined. The number of radial bins is determined with a modified version of the binning strategy from M. Zemp et al. ([2011](#)):

$$n_b = \begin{cases} \dfrac{n_p}{1000} + 10 & \text{for } n_p < 10^4, \\ 20 \cdot (\log_{10} n_p - 3)^2 & \text{for } n_p \geqslant 10^4, \end{cases} \quad (4)$$

where $n_b$ represents the number of radial bins and $n_p$ denotes the total count of DM or star particles contained within the maximum stellar radius. We choose to require a minimum of 5000 star particles, which enforces at least 15 radial bins and a minimum of 300 star particles per bin. By selecting galaxies with at least 5000 star particles, we ensure a minimum of 54,000 DM particles, with at least 8000 located within the stellar region.

This adaptive binning function extends M. Zemp et al. ([2011](#))'s approach by incorporating a steeper scaling relationship to increase radial resolution when particle counts are high ($n_p \sim 10^7$) while reducing bin numbers for lower particle densities ($n_p \sim 10^4$).

The requirement for the number of particles is the primary limiting factor in our selection of galaxies from Marvel and DCJL suites. In the Massive Dwarfs, satellites of isolated field dwarf galaxies were not considered, and the isolated field dwarfs are well above the minimum particle threshold. The lowest-mass galaxies we probe for each suite is $1.1 \times 10^6 \, M_\odot$, $2.5 \times 10^7 \, M_\odot$, and $9.9 \times 10^7 \, M_\odot$ for Marvel, DCJL, and Massive dwarfs, respectively. However, these requirements restrict our sample to higher stellar masses ($M_* \gtrsim 10^6 \, M_\odot$) compared to the faintest observed galaxies and many smaller galaxies within our simulation suites.

We focus our analysis on measurements taken at twice the effective radius (2 $R_{eff}$), which we determine by fitting the logarithmic form of the Sérsic profile as presented in Equation (4) of N. Caon et al. ([1993](#)) to the V-band surface brightness profile in a face-on orientation calculated with `pynbody` (A. Pontzen et al. [2013](#)). The face-on orientation is determined by aligning the viewing angle perpendicular to the disk plane, which is defined by the total angular momentum vector of the gas component within a 5 kpc radius (A. Pontzen et al. [2013](#)). We selected this radius because it falls within the region where SIDM shape differences may be detectable. For an ideal Sérsic profile, $R_{eff}$ is equivalent to the half-light radius ($R_{1/2}$), though in practice these values may differ slightly. $R_{eff}$ naturally scales with system size, allowing for consistent comparisons across our sample. While we present our primary results at 2 $R_{eff}$, we analyze shapes at $R_{eff}$ to connect with previous work and verify radial trends (see Appendix [D](#)).

### 2.4. Resolution

As described in Section [2.3](#), to ensure sufficient resolution for shape measurements, we require at least 15 bins and 5000 star and DM particles. This strict requirement exclusively selects galaxies that are well resolved both spatially and in mass.

Previous studies in the Marvel and DCJL simulations have demonstrated that the spatial resolution limits for these simulations is 0.25 kpc using criteria from C. Power et al. ([2003](#)). More conservative limits from A. D. Ludlow et al. ([2020](#)) constrain the resolved spatial scales at 0.45 kpc in Marvel, and 0.8 kpc in DCJL (C. L. Riggs et al. [2024](#)). In another study, C. A. Vera-Ciro et al. ([2014](#)) find that DM shapes are accurate up to ∼8% at the resolution limit from C. Power et al. ([2003](#)). Our galaxy selection criteria results in galaxies with 2 $R_{eff}$ greater than any of these limits, ensuring our shape measurements are only taken in regions that are well resolved.

### 2.5. Disk Identification

To investigate whether stellar shapes differ from DM shapes based on whether a galaxy has a disk, we classified galaxies by their intrinsic shape. Galaxies that are thin ($S_* < 0.4$) when viewed edge on and circular when viewed face on ($Q_* > 0.65$) at 2 $R_{eff}$ are considered disky, while galaxies that do not meet these criteria are considered nondisky. We confirm that this behaves as expected through visual inspection of all our galaxies, using four images: one face-on, and three edge-on orientations separated by 120°. We show three example galaxies in Figure [1](#) showing the face-on orientation and one of the side-on orientations. Our final sample contains 34 disky galaxies and 46 nondisky galaxies.

We acknowledge that our results depend on these specific threshold values; however, only 14% of galaxies fall within a ±0.05 margin of either of these boundaries, and such modifications do not significantly impact our interpretation.

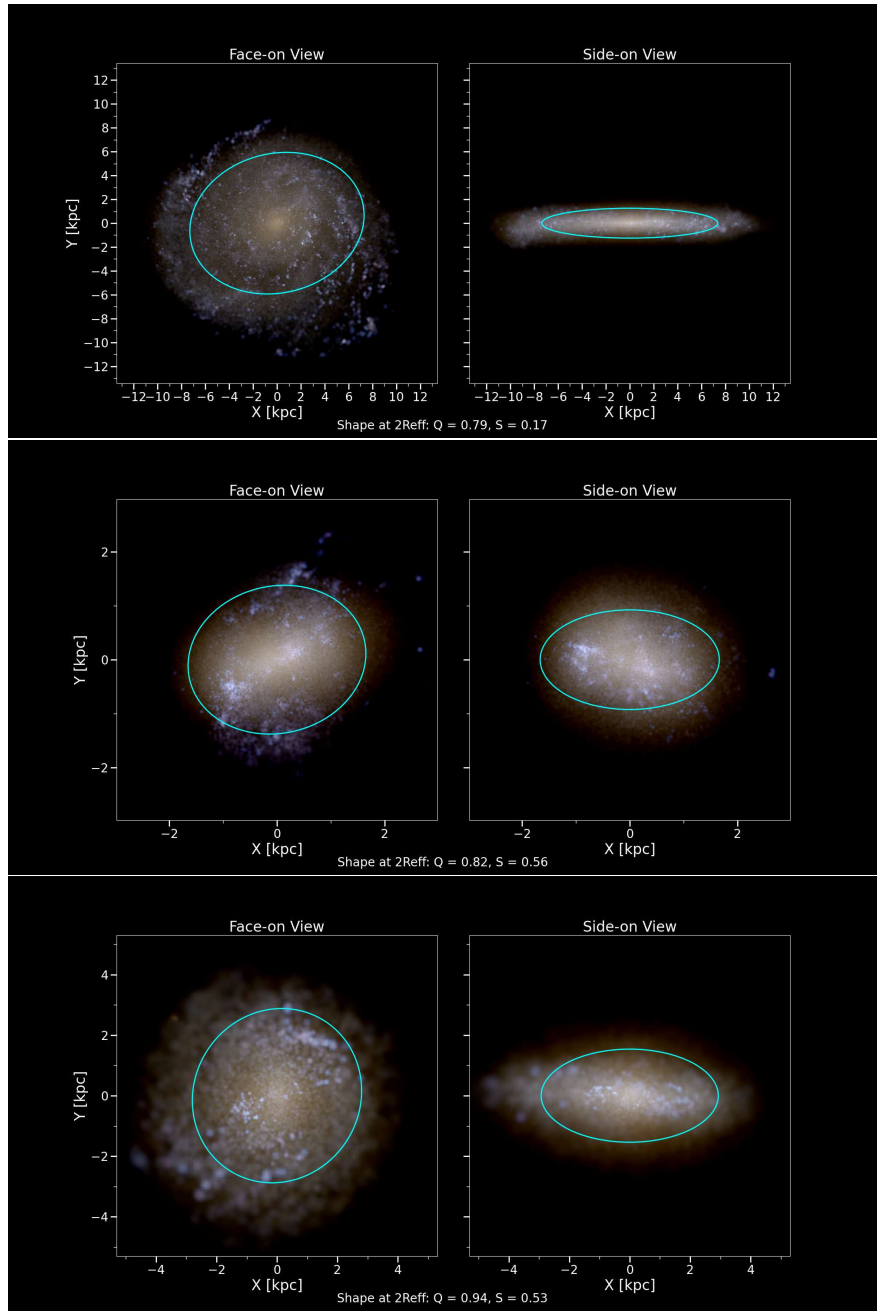

**Figure 1.** Simulated red–green–blue images in the *i*, *V*, and *u* bands generated using `pynbody.plot.stars.render`. Left panels show face-on orientations, while right panels show edge-on orientations. The cyan projected ellipse indicates shape and scale at 2 $R_{\rm eff}$. Following the disk classification criteria defined in Section 2.5, we present three representative cases: (top) a clear disky galaxy; (middle) a nondisky galaxy; and (bottom) a galaxy with visual thick disk characteristics that does not satisfy our classification criteria.

Moreover, our classification method demonstrates consistency, aligning with visual classifications for 95% of galaxies and agreeing with kinematic measurements—based on angular momentum—for 88% of the sample, as described in more detail in Appendix B.

### 2.6. Merger Identification

We identify mergers using "The Agile Numerical Galaxy Organization System" (A. Pontzen & M. Tremmel 2018), which employs `pynbody`'s (A. Pontzen et al. 2013) `bridge` function to track shared DM particles across time steps. We identify the merger ratio as the mass (DM + baryons) of the major progenitors divided by the combined mass of all progenitors. We do not expect that mergers with ratios $\gtrsim 5$ have sufficient mass to dramatically impact the 3D shape of a much more massive progenitor (C. A. Vera-Ciro et al. 2014).

## 3. Results

### 3.1. Observational Validation of Simulated Dwarf Galaxy Shapes

Our simulated sample successfully recreates distributions of shape trends captured from observations. First, we examine the mass dependence of galaxy shapes to validate our simulation results against observational data. We measure local shapes of galaxies by fitting ellipsoid shells, as described in Section 2.3. The axis ratios $Q = B/A$ and $S = C/A$ provide a quantitative measure of the galaxy's (or DM halo's) shape: $Q = 1$ indicates circular symmetry in the $B$–$A$ plane, which for galaxies with a stellar disk corresponds to the disk plane itself. Lower values of $Q$ indicate elongation in the disk plane, while small values of $S$ indicate increasing vertical flattening of the galaxy perpendicular to the disk plane.

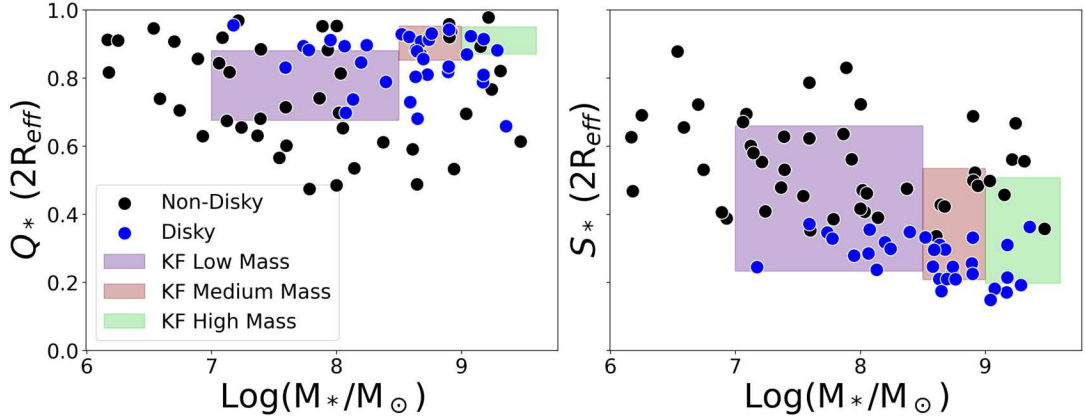

**Figure 2.** Axis ratios of stars as a function of stellar mass ($[\log(M_*/M_\odot)]$). The left panel shows $Q_*$, where $Q$ (intermediate-to-major axis ratio) indicates elongation along the major axis, while the right panel shows $S_*$, where $S$ (minor-to-major axis ratio) indicates the degree of flattening. Blue points represent disky galaxies, while black represents nondisky. Shaded regions show maximum a posteriori estimates with $1\sigma$ standard deviations from E. Kado-Fong et al. (2020) for different mass bins: low mass (indigo, $7.0 \leqslant \log(M_*/M_\odot) \leqslant 8.5$), medium mass (dark red, $8.5 \leqslant \log(M_*/M_\odot) \leqslant 9.0$), and high mass (lime green, $9.0 \leqslant \log(M_*/M_\odot) \leqslant 9.6$).

Figure 2 shows the relationship between stellar mass and axis ratio ($Q_*$ and $S_*$) measured at $2\,R_{\rm eff}$ for both disky and nondisky galaxies. The average minor-to-major axis ratio ($\bar{S}$) differs significantly between these populations, with disky galaxies showing $\bar{S}_* = 0.26 \pm 0.06$ and nondisky galaxies having $\bar{S}_* = 0.54 \pm 0.13$. The intermediate-to-major axis ratio $Q_*$ shows only a weak mass dependence on stellar mass (slope = $-0.03 \pm 0.02$). The minor-to-major axis ratio $S_*$ exhibits more of a correlation with stellar mass, having slopes of $-0.043 \pm 0.02$ for nondisky and $-0.053 \pm 0.02$ for disky galaxies. This trend in $S_*$ indicates that more massive systems become increasingly flattened, even in the absence of a clear stellar disk.

In our sample, we find that stellar disks form predominantly above $M_* \gtrsim 10^{7.5}\,M_\odot$. Below this mass ($10^6 \lesssim M_*/M_\odot \lesssim 10^{7.5}$), our simulated galaxies do not effectively form disks. However, not every galaxy above this threshold is a disk; we find that 54% are disky and at a higher-mass threshold of $10^{8.5}\,M_\odot$ the fraction of disky galaxies increases to 64%. We further discuss feedback and disk formation in Section 4.2.

Our simulated galaxy shapes show agreement with multiple observational studies across comparable mass ranges. R. Sánchez-Janssen et al. (2010) find that disk thickness reaches a minimum ($S \approx 0.2$) around $M_* \sim 2 \times 10^9\,M_\odot$, consistent with our disky population, though our sample's upper mass limit prevents us from confirming their observed increase in thickness at higher masses ($M_* \gtrsim 10^{10}\,M_\odot$). At lower masses, our results align with both S. Roychowdhury et al. (2013) and E. Kado-Fong et al. (2020), who observe thicker stellar distributions ($S \approx 0.4$) in intermediate-mass dwarfs ($M_* \approx 10^7$–$10^8\,M_\odot$) compared to more massive dwarfs ($M_* \approx 10^9\,M_\odot$, $S \approx 0.2$–$0.3$). This mass-dependent transition from thicker, more spheroidal shapes to flatter, disk-dominated shapes reflects the increasing prominence of stellar disks in more massive systems, a structural trend successfully captured by our simulations.

More specifically, we compare our shape measurements to those of E. Kado-Fong et al. (2020) across three stellar mass bins: $10^{7.0}$–$10^{8.5}\,M_\odot$ (low), $10^{8.5}$–$10^{9.0}\,M_\odot$ (medium), and $10^{9.0}$–$10^{9.6}\,M_\odot$ (high). In Figure 2, these bins are represented by indigo, dark red, and lime green shaded regions, respectively, showing the $1\sigma$ confidence intervals of the maximum a posteriori estimates for their dwarf galaxy sample. The minor-to-major axis ratios ($S_*$) show excellent agreement across all mass ranges, with differences of $\Delta S_* \leqslant 0.05$ between our measurements and their observational results. The strongest agreement is found in the high-mass bin ($\Delta S_* = 0.01$), with our measured $S_* = 0.36 \pm 0.17$ comparing well to their value of $S_* = 0.35 \pm 0.16$. The intermediate-to-major axis ratios ($Q_*$) show increasing deviation with mass, from $\Delta Q_* = 0.01$ in the low-mass bin to $\Delta Q_* = 0.1$ in the high-mass bin. Notably, our measurements exhibit larger variations in $Q_*$, with standard deviations larger than those reported by E. Kado-Fong et al. (2020). We find $\sigma_{Q_*} = (0.11, 0.13, 0.14)$ compared to their values (0.04, 0.05, and 0.10) for high-, medium-, and low-mass bins, respectively. This increased scatter could result from our measurement of simulated intrinsic 3D shapes compared to their de-projected observations. Additionally, the stellar mass observed within our simulations may not necessarily align with the stellar masses from E. Kado-Fong et al. (2020) as they are estimated from $B$-band brightness measurements and may not match simulations (e.g., M. A. C. de los Reyes et al. 2024).

### 3.2. Stellar versus Dark Matter Shapes

Next, we investigate the shape of both DM and SM (the galaxies), to see if stellar shape (galaxy shape) can reliably be used to infer DM shape. We start by examining the dwarf population as a whole in Figure 3, and then examine individual galaxies in Figure 4. We find that the stellar shapes are the most reliable tracers of DM shapes in galaxies that do not have a stellar disk, and that the DM and SM shapes become more similar as galaxy mass decreases.

Figure 3 presents a comparison of the axis ratios $Q$ and $S$ at $2\,R_{\rm eff}$. The axis ratios of stars are denoted by the star symbol,

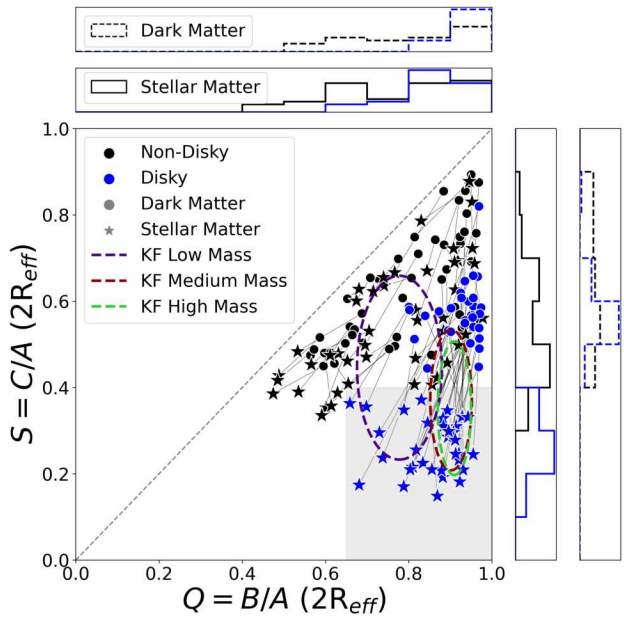

**Figure 3.** Axis ratio measurements for DM and stars at twice the effective radius (2 $R_{\rm eff}$). The main plot shows the minor-to-major axis ratio ($S = C/A$) vs. the intermediate-to-major axis ratio ($Q = B/A$) for DM (circles) and stellar components (stars). Each galaxy is linked to its halo with a solid gray line. The gray dashed 1:1 line represents perfectly prolate objects where $B = C$. The region above this is empty as $C < B < A$ by definition. Blue denotes disky galaxies while black represents nondisky galaxies. Note for $S$, $Q \sim 1$ represents spherical objects that are not well described by prolate or oblate. Dashed ellipses show the maximum a posteriori estimates with $1\sigma$ standard deviations from E. Kado-Fong et al. (2020) for different mass bins: low mass (indigo, $7.0 \leqslant \log(M_*/M_\odot) \leqslant 8.5$), medium mass (dark red, $8.5 \leqslant \log(M_*/M_\odot) \leqslant 9.0$), and high mass (lime green, $9.0 \leqslant \log(M_*/M_\odot) \leqslant 9.6$). Histograms of $Q$ and $S$ distributions are displayed in the top and right panels, respectively, where stars are represented by solid lines and DM represented by dashed lines. The shaded gray region represents the disky selection criteria of $S_* < 0.4$ and $Q_* > 0.65$. These measurements reveal that nondisky galaxies tend to have similar shapes in both their DM and stellar distributions, while disky galaxies show a clear difference: their DM halos are more spherical while their stellar components form flattened disk-like structures.

while the axis ratios of DM are denoted by circles. Lines connect the SM and DM shapes for individual galaxies. The accompanying histograms illustrate the number counts for axis ratio in DM (dashed lines) or in stars (solid lines). The $Q$ distributions of DM and SM generally overlap, with a range of values from 0.56 to 0.98, although stellar shapes extend all the way down to $Q = 0.47$. In general, the SM has a tail in the distribution that extends to lower $Q$, though we show in Figure 4 that the difference is small for any individual galaxy.

Galaxies with a stellar disk have stellar shapes that occupy the lower region of Figure 3. The stellar axis ratios, $S_*$, are lower than their corresponding DM axis ratios, $S_{\rm DM}$, due to the presence of the stellar disk ($\bar{S}_* = 0.26 \pm 0.06$, $\bar{S}_{\rm DM} = 0.57 \pm 0.07$). On the other hand, the $S$ values of nondisky galaxies are more similar to each other. Overall, the distribution of DM and stellar shape are similar for galaxies without a notable stellar disk, with over 75% of values concentrated near the $Q = S$ line representing prolate objects ($S \approx (Q - 0.15) \pm 0.15$).

A two-sample Kolmogorov–Smirnov test confirms the above visual inspections. For nondisky galaxies, the $Q$ distributions between stars and DM are statistically indistinguishable ($p = 0.49$), while their $S$ distributions show modest differences ($p = 0.015$). In contrast, disky galaxies show significant differences in both parameters ($p < 10^{-5}$ for $Q$ and $p < 10^{-19}$ for $S$), with $S$ showing the maximum possible difference (KS statistic $= 1.0$). These statistical tests support our finding that stellar shapes are reliable tracers of DM shapes in galaxies that lack a stellar disk.

We compare our simulated shape distributions to observational measurements from E. Kado-Fong et al. (2020), who derived intrinsic shapes for observed galaxies in three stellar mass bins. Our disky galaxies show good agreement with their intermediate- and high-mass bins ($M_* > 10^{8.5} M_\odot$), though our sample includes a notable population with lower $Q$ values than found in their observations. The observational low-mass bin partially overlaps with both our disky and nondisky populations, but is centered in a relatively unpopulated region of our parameter space. While our simulations predict nondisky galaxies across a range of shapes, including both less and more spherical systems than is observed, this discrepancy may reflect limitations in observational shape de-projection techniques rather than physical differences. Specifically, the de-projection methods may struggle to fully capture more complex and extended stellar distributions, potentially biasing the recovered shape distribution.

To explore whether the central stellar mass distribution influences the DM shape, Figure 4 shows stellar mass plotted against the ratios $Q_{\rm DM}/Q_*$ and $S_{\rm DM}/S_*$. Disky galaxies exhibit the highest $S_{\rm DM}/S_*$ ratios, averaging $2.92 \pm 0.60$, with this difference increasing with stellar mass (slope $= 0.56 \pm 0.02$, residuals $1\sigma = 0.53$). However, their $Q_{\rm DM}/Q_*$ shows no significant mass dependence (slope $= 0.017 \pm 0.032$). Nondisky galaxies display modest trends, with $S_{\rm DM}/S_*$ and $Q_{\rm DM}/Q_*$ showing weak mass dependence (slopes $= 0.11 \pm 0.02$ and $0.032 \pm 0.016$, respectively). Combining both samples together still yields a weakly increasing of $Q_{\rm DM}/Q_*$ with mass (slope $= 0.033 \pm 0.013$) The stronger trend in $S_{\rm DM}/S_*$ with stellar mass again suggests that there is a general trend toward flattening of the SM, even in nondisky galaxies, as was suggested in Figure 2.

The increased flattening of $S_*$ with mass is in line with expectations from previous works examining galaxies at larger stellar masses (A. Tenneti et al. 2014; M. Velliscig et al. 2015; K. T. E. Chua et al. 2022), though L. M. Valenzuela et al. (2024) find no such dependence. These works have also found that $Q$ is systematically larger in DM compared to SM in more massive galaxies, but we are working with simulated galaxies at lower masses which do not show as strong of a trend in $Q$. At lower stellar mass ranges ($10^5$–$10^7 M_\odot$), M. D. A. Orkney et al. (2023) found that gas-poor systems maintain prolate DM halos even with baryonic physics included, while gas-rich ones become rounder and more oblate within 10 $R_{1/2}$. These trends are similar to the mass trends that we find for disky and nondisky systems.

Finally, we find no noticeable differences in shape trends when considering environment: specifically field versus satellite dwarf. In Figure 13 of Appendix C we recreate Figure 4 but highlight satellites. Unsurprisingly, the majority of satellites are nondisky (11/15).

In summary, we find that the shapes of DM and SM trace each other most reliably in nondisky galaxies, which dominate our galaxy sample at lower stellar masses $M_* \lesssim 10^{7.5} M_\odot$. Meanwhile, galaxies with stellar disks show greater differences between their DM and stellar shapes, with this difference

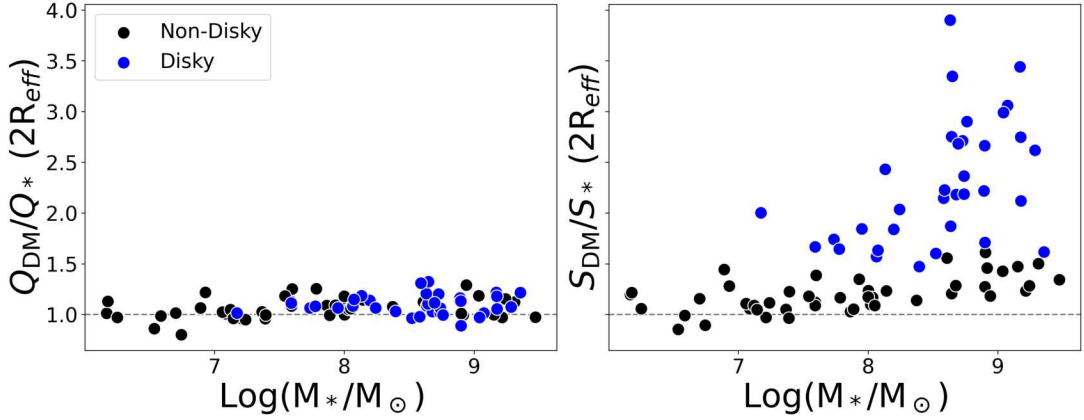

**Figure 4.** Stellar mass ([$\log(M_*/M_\odot)$]) vs. relative changes in shape between DM and stellar components. The left panel shows $Q_{\rm DM}/Q_*$, where $Q$ (intermediate-to-major axis ratio) indicates elongation along the major axis, while the right panel shows $S_{\rm DM}/S_*$, where $S$ (minor-to-major axis ratio) indicates the degree of flattening. Blue points represent disky galaxies, while black represents nondisky. The dashed horizontal line shows when DM and stars have equal shapes ($Q_{\rm DM} = Q_*$ or $S_{\rm DM} = S_*$). The plot reveals that while the elongation of DM and stellar components remains similar across galaxy masses (left panel), more massive disky galaxies show increasingly flattened stellar distributions compared to their DM halos (right panel), reflecting the formation of prominent stellar disks in these systems.

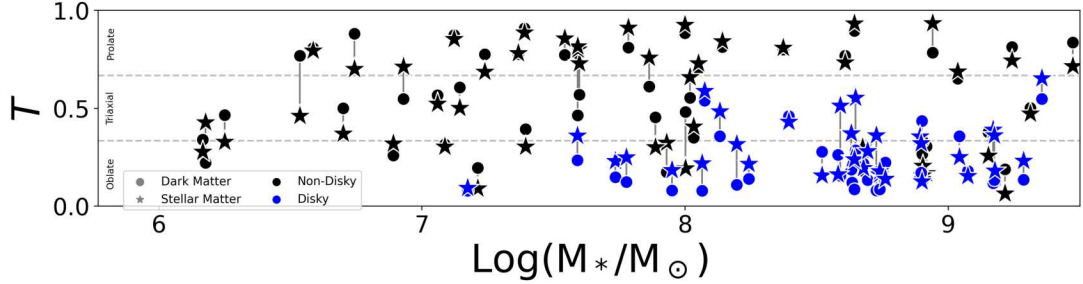

**Figure 5.** Triaxiality, $T = (1 - Q^2)/(1 - S^2)$, for our galaxies as a function of stellar mass ([$\log(M_*/M_\odot)$]). Vertical lines link the $T$ values for DM (circles) and SM (stars) for each galaxy. Galaxies with pronounced stellar disks are colored blue, those without are black. The measurements reveal that lower-mass galaxies tend to be more elongated (prolate) in both their stellar and DM distributions, while higher-mass disky galaxies typically have flattened disk-like (oblate) stellar distributions. Most galaxies show similar shapes in their DM and stellar components, though the DM tends to be slightly more elongated in the lowest-mass systems.

increasing as stellar mass increases (corresponding to more disky galaxies).

### 3.3. Triaxiality

It is common for 3D shapes to be quantified through their triaxiality parameter, $T$, as defined in M. Franx et al. (1991):

$$T = \frac{1 - Q^2}{1 - S^2}. \quad (5)$$

A shape is considered oblate when $0 < T < \frac{1}{3}$, triaxial when $\frac{1}{3} < T < \frac{2}{3}$, and prolate when $\frac{2}{3} < T < 1$. To aid the reader, the distribution of triaxiality values on a $Q$ versus $S$ plot is shown in Figure 11 (Appendix A).

In order to investigate whether stellar triaxiality serves as a reliable proxy for DM halo triaxiality, we examine the relationship as a function of stellar mass in Figure 5. The plot shows both DM (circles) and SM (stars) triaxiality measurements, with vertical lines connecting each galaxy to its corresponding DM halo. Disky galaxies cluster in the oblate region ($T \lesssim 1/3$; mean $T_{\rm DM} = 0.20 \pm 0.13$, $T_* = 0.29 \pm 0.14$). Nondisky galaxies tend toward triaxial and prolate values ($T \gtrsim 2/3$; mean $T_{\rm DM} = 0.57 \pm 0.23$, $T_* = 0.56 \pm 0.26$). There are some outliers from these more dominant trends. Previous works have investigated the origin of outliers in detail, attributing their deviations to variations in DM halo assembly history at the specific measurement radii (see B. Moore et al. 2004; M. D. A. Orkney et al. 2023). The 2 $R_{\rm eff}$ measurements used in this work range from 0.6 to 9.5 kpc over our sample.

A direct comparison between DM and stellar triaxiality is shown in Figure 6. Nondisky galaxies exhibit a strong correlation between $T_*$ and $T_{\rm DM}$, with a near-unity slope of $0.99 \pm 0.08$ and $R^2 = 0.77$. In contrast, disky galaxies show a weaker correlation (slope = $0.54 \pm 0.16$, $R^2 = 0.25$) and deviate significantly from the one-to-one relationship. This deviation arises because DM in disky galaxies tends to be more oblate than their stellar components, primarily due to larger $Q$ values. The effect of $Q$ on $T$ dominates over the influence of smaller $S$ values when $Q \sim 1$ (see also Figure 11). The absolute difference between stellar and DM triaxiality is shown in Figure 7. We find relatively small differences for most of our sample, with only 20 out of 80 dwarfs showing differences greater than 0.15.

Despite disky galaxies having large differences in $S$ ($S_{\rm DM}/S_* = 2.92$), their triaxiality values remain similar ($|T_{\rm DM} - T_*| \lesssim 0.15$). This occurs because the shape changes primarily follow triaxiality contours (see Appendix A). This highlights a limitation of the triaxiality parameter: significant changes in $S$ can be masked along a triaxiality contour, while for more spherical shapes, small changes in $Q$ and $S$ can result in large triaxiality variations.

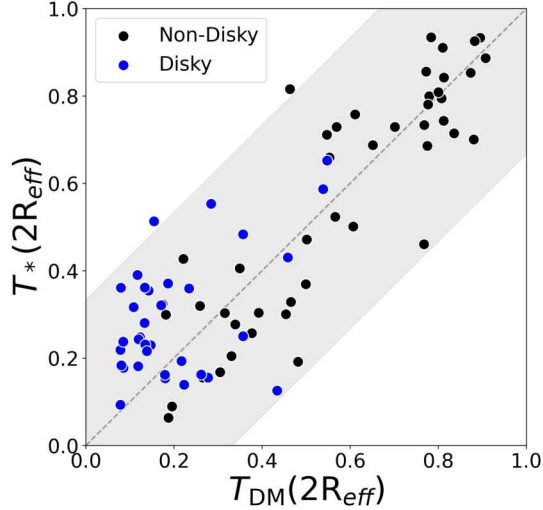

**Figure 6.** Triaxiality of the DM ($T_{DM}$) vs. the triaxiality of the SM ($T_*$) at twice the effective radius (2 $R_{eff}$). Disky galaxies are colored in blue, while nondisky are black. The dashed line shows $T_* = T_{DM}$ The shaded gray region represents the vicinity of the dashed line, defined by $T_* = T_{DM} + c$ where $-1/3 \leqslant c \leqslant 1/3$. This comparison reveals that DM and stellar triaxialities generally track each other, though in the most oblate and disky systems, the SM is less oblate than the DM.

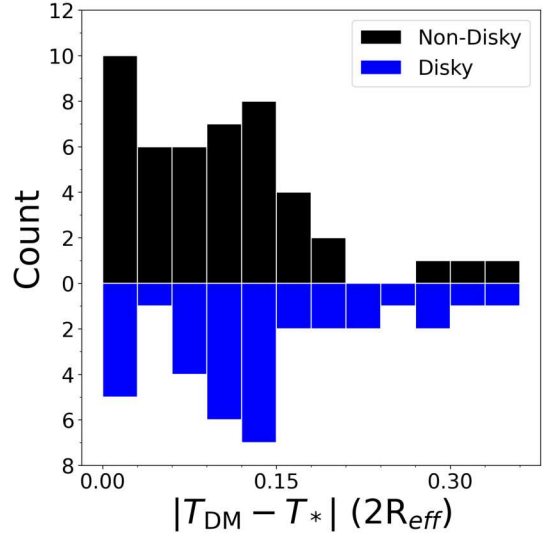

**Figure 7.** Histogram showing the absolute difference between stellar and DM triaxiality ($|T_* - T_{DM}|$) at twice the effective radius (2 $R_{eff}$). Blue represents disky galaxies while black represents nondisky galaxies. The distribution reveals that most galaxies show relatively small differences between their stellar and DM triaxiality values, though disky galaxies tend to show slightly larger differences compared to their nondisky counterparts.

Our results are broadly consistent with observational studies. E. Kado-Fong et al. (2020) find in their lower-mass sample ($10^7 - 10^{8.5}$ $M_\odot$) more triaxial values ($1/3 \lesssim T \lesssim 2/3$). However, their higher-mass samples ($10^{8.5} - 10^9$ $M_\odot$) tend toward more oblate shapes ($T \lesssim 1/3$), similar to our simulated dwarfs. At classical dwarf stellar masses ($10^5 - 10^7$ $M_\odot$), W. L. Xu & L. Randall (2020)'s FIRE simulations found generally prolate values for stars at $R_{1/2}$ ($T \approx 2/3$), similar to our measurements at 1 $R_{eff}$ (average $\bar{T}_* = 0.59 \pm 0.23$; see Figure 15), and our disky sample does include more oblate values for shapes that W. L. Xu & L. Randall (2020) found to be missing in their FIRE simulations at similar masses compared to their observations of LG dwarfs.

Finally, we note that the interpretation of triaxiality measurements requires careful consideration of error propagation. Small uncertainties in the axis ratios can lead to significant uncertainties in $T$, particularly when $S$ approaches one. The propagated error in $T$ can be expressed as

$$\sigma_T = \sqrt{\left(\frac{-2Q}{1-S^2}\sigma_Q\right)^2 + \left(\frac{2S(1-Q^2)}{(1-S^2)^2}\sigma_S\right)^2}, \quad (6)$$

where $\sigma_Q$ and $\sigma_S$ represent the uncertainties in the axis ratios. When comparing shapes within individual simulations, where convergence criteria constrain uncertainties to <0.01 in $Q$ and $S$, the propagated error in $T$ remains manageable ($\sigma_T \lesssim 0.06$) except when $S \gtrsim 0.8$. When comparing across different simulations or with observations, typical uncertainties in axis ratios may reach ~0.1 (e.g., M. D. A. Orkney et al. 2023). While this uncertainty would be significant for individual galaxies, particularly those approaching spherical shapes, its impact is reduced when comparing larger samples like our sample of 80 galaxies, as random errors are expected to average out. This suggests that triaxiality comparisons are most reliable when (1) examining systems that are distinctly nonspherical ($S \lesssim 0.8$), (2) comparing measurements within the same simulation, or (3) when analyzing trends across larger samples where individual measurement uncertainties become less significant.

### 3.4. Alignment of Shapes Between Dark Matter and Stars

The distributions of axis ratios, $S$ and $Q$, between DM and SM have overlap, especially in nondisky galaxies, but to see if they both truly trace the same underlying total matter distributions, the orientations of the measured 3D shapes must also align. We find that the orientation of shapes between DM and SM are generally aligned, especially for the $C$ axis, which is typically aligned with the total angular momentum of a gaseous disk when present (V. P. Debattista et al. 2008; M. D. A. Orkney et al. 2023).

We investigate the orientations of fitted ellipsoid shells between DM and SM. When any two (or three) axes of an ellipsoid are approximately equal in length, their orientations become physically less meaningful, with measured orientations appearing randomly distributed even when the underlying matter distributions are well aligned. This degeneracy occurs in oblate systems ($B \approx A$), prolate systems ($B \approx C$), or spherical systems (all axes nearly equal).

An axis is considered well aligned when the orientation between DM and SM ellipsoid axes differ by less than 10°. The misalignment angles between axes are given by

$$\Delta\alpha = \arccos(|\boldsymbol{e}_{i,*} \cdot \boldsymbol{e}_{i,DM}|), \quad (7)$$

where $\boldsymbol{e}_{i,*}$ and $\boldsymbol{e}_{i,DM}$ are the corresponding column vectors of rotation matrices $\boldsymbol{R}_*$ and $\boldsymbol{R}_{DM}$ (representing SM and DM orientations at 2 $R_{eff}$), with $i \in \{x, y, z\}$ corresponding to the major ($A$), intermediate ($B$), and minor ($C$) axes respectively. Due to axial symmetry of ellipsoids, the misalignment angle cannot be greater than 90°.

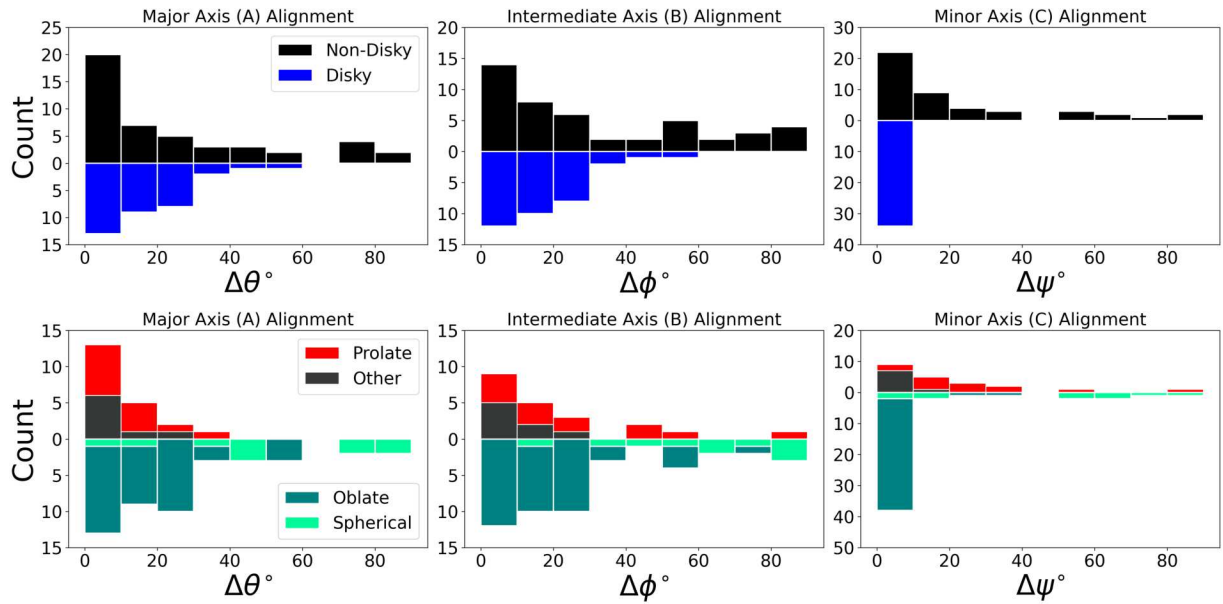

**Figure 8.** Distribution of misalignment angles between DM and stellar components measured at 2 $R_{\rm eff}$, showing the alignment of the major ($A$, left panels), intermediate ($B$, middle panels), and minor ($C$, right panels) axes. Top panels: histograms show the number of galaxies in 10° bins, with disky galaxies in blue and nondisky galaxies in black. The minor $C$ axis (right panel) shows the strongest alignment, particularly for disky galaxies where this axis typically aligns with the total angular momentum. Bottom panels: the same distributions categorized by system shape characteristics, with systems classified based on either their DM or stellar component being prolate (red), oblate (teal), spherical ($S > 0.8$, mint green), or other (dark gray). This classification reveals how the alignment of each axis depends on the underlying shape of the galaxy.

Figure 8 presents the distribution of misalignment angles between the DM and SM components for each principal axis. We find distinct alignment patterns, with the strongest correlation along the $C$ axis (60/80 galaxies show strong alignment). This alignment is particularly pronounced in disky galaxies, where all such systems demonstrate strong alignment, compared to only 50% of nondisky galaxies. The $A$ and $B$ axes exhibit weaker alignment patterns in the disky sample, expected as many disky galaxies have intermediate axes only slightly shorter than their major axes. In nondisky galaxies, which are often prolate, we find significant $A$ axis alignment, with weaker alignments in the $B$ and $C$ axes.

We quantify the role of shape degeneracy in these alignments by categorizing our sample into four exclusive categories: oblate ($T < 1/3$), prolate ($T > 2/3$), spherical ($S > 0.8$), or other. A galaxy is classified in one of these categories if either its DM or SM component meets the corresponding criterion. Figure 8 (bottom panels) demonstrates that the vast majority of the sample shows alignment patterns consistent with their underlying shapes. The apparent misalignments can largely be explained by near-degeneracies in axis lengths: oblate galaxies show strong $C$ axis alignment (38/41 galaxies) but weaker correspondence in $A$ and $B$ axes, while prolate galaxies demonstrate strong $A$ axis alignment (13/21 galaxies). Spherical galaxies (10) account for most cases showing misalignment across all three axes, as expected given their nearly equal axis lengths. The remaining galaxies in our sample show good alignment across all axes (5/8). In total, we conclude that 82% of our sample is aligned according to our expectations from shapes.

Our results for alignment stand in contrast with previous work by A. Tenneti et al. (2014) and L. M. Valenzuela et al. (2024), though they examined more massive galaxies than we do. A. Tenneti et al. (2014) found dependencies on the alignment between DM and SM shape orientations, such that the mean misalignments decrease from 34° to 14° as mass increases from $\sim 10^{11}$ to $> 10^{13.0}$ $h^{-1}$ $M_\odot$). Similarly, L. M. Valenzuela et al. (2024) found better alignment in late-type galaxies (77% well aligned) compared to early type (49% well aligned). However, A. Tenneti et al. (2014) focused on alignments for only the major axis while computing a shape for the entire galaxy. Meanwhile, L. M. Valenzuela et al. (2024) only examined the major axis at 3 $R_{1/2}$. The fact that we find stronger alignment might be due to our consideration of additional axes and shape categories, prolate and oblate. A. Tenneti et al. (2014) did consider spherical shapes, but used a threshold to define spherical shapes that is higher than ours ($S > 0.95$ versus $S > 0.8$).

### 3.5. Mergers

Mergers can significantly alter galaxy shapes through tidal distortions, disruption of stellar orbits, gas dynamics, and redistribution of DM (L. Wang et al. 2019). We investigated the shapes of both DM and SM before, during and postmerger, in order to rule them out as a possible source of difference between DM and stellar shape. We find that mergers do not significantly impact our results.

Since we are focused on measurements at redshift zero, we only consider previous time steps that could impact 3D shape measurements at redshift zero. We estimate how long it takes for a galaxy to return to equilibrium after a merger using dynamical times calculated at the maximum extent of the galaxy. We find in our sample, dynamical times range from 20 to 150 Myr. Note that dynamical times at $R_{\rm eff}$ will be even shorter. Each output time step represents 215 Myr, so we expect that measuring the last three output time steps ($\sim 600$ Myr) of our simulations is sufficient to ensure that the regions where stars are located in merging galaxies have had time to reach equilibrium from any previous mergers. We find in our simulations two galaxies with recent mergers that have merger ratios of $\sim 5$, with ratios 4.4 and 5.6, which by some standards may not be considered major. During a recent merger, stellar and DM shapes have limited movement. Figure 9 shows neither $Q$ nor $S$ vary more than $\sim 0.05$ across the total time period ($z = 0$–$0.04$). This result would hint at the idea that even mergers of similar mass do not seem to dramatically rearrange the central density structure of DM halos (e.g., B. Moore et al. 2004). With only two merger events with merger ratios $\sim 5$ in the last 600 Myr, merger events are also rare enough in our sample to not skew our results. We conclude that for dwarf galaxies with merger ratios $\gtrsim 4$, mergers do not have a significant effect on the distribution of stars, which in our study is the shape of SM. However, given that our sample lacks mergers with merger ratios approaching one, we can only constrain the impact of mergers with merger ratios $> 4$ on the shapes of both DM and SM.

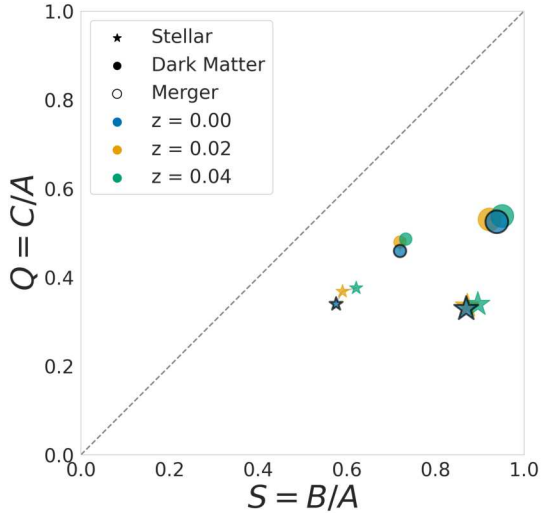

**Figure 9.** For galaxies that have had a recent merger, we show the evolution of $Q = B/A$ and $S = C/A$ for the last three time steps at $2\,R_{\rm eff}$. DM (circles) and stars (asterisks) are linked across time, showing their recent evolution. Points that are outlined in black mark the time step that the largest recent merger occured. The point size is scaled by largest recent merger for each galaxy over the measured time period. The latest time at $z = 0$ is shown in green, and previous times at $z = 0.02$ in orange, and $z = 0.04$ in blue. The trajectories reveal that mergers typically cause modest changes in galaxy shapes, with axis ratios varying by no more than $\sim 0.05$ between time steps, suggesting that mergers with ratios $> 4$ do not dramatically alter the overall shape of either the DM or stellar distributions.

## 4. Discussion

### 4.1. Influence of Baryons on the Shape of Dark Matter

When focusing efforts on studying DM models, ideally, we would want to isolate the effects of baryons on the shapes of DM as much as possible. By comparing the shapes of DM between galaxies with and without stellar disks, our data imply that nondisky galaxies have decreased baryonic influence that impacts the shape of DM, reinforcing that these galaxies are more ideal to study DM models.

Comparing the average DM shapes between nondisky and disky galaxies, we find that disky galaxies are more circular ($Q_{\rm DM}$ increases from 0.80 to 0.93), flatter ($S_{\rm DM}$ decreases from 0.64 to 0.57), and significantly more oblate ($T_{\rm DM}$ decreases from 0.57 to 0.20). Therefore, the presence of a stellar disk appears to influence the shape of DM in our simulated galaxy sample at $2\,R_{\rm eff}$. In DMO-only simulations of low-mass halos ($V_{\rm max} \approx 50\ {\rm km\ s^{-1}}$ or $M_* \approx 2 \times 10^9\ M_\odot$) C. A. Vera-Ciro et al. (2014) find values for classical satellites of the Milky Way, $S \approx 0.60$ and $Q \approx 0.75$ at radii near 1 kpc, similar to our nondisky dwarfs. This agreement between DMO and full physics simulations suggests that in these mass ranges, baryonic effects minimally impact the overall shape of the DM distribution. Our simulations suggest a mass threshold of $\sim 10^{7.5}\ M_\odot$ where baryonic effects begin to impact DM shapes through disk formation. This finding extends previous analyses that primarily focused on higher-mass ranges, such as the work by K. T. E. Chua et al. (2019) who examined halos from $4 \times 10^8$ to $10^{12}\ M_\odot$ and found mass-dependent baryonic effects. By probing lower-mass scales, we have identified a transition point where disk formation starts to alter DM distributions within our simulations.

### 4.2. Subgrid Supernova Feedback

We combine two different SN feedback models to increase our sample size, confident that this approach does not significantly impact our shape measurements. As mentioned in Section 2, Marvel and DCJL, which contribute satellite systems in this study, both implement the blastwave SN feedback model, while Massive Dwarfs implements superbubble and were selected to contain field dwarfs.

Previous work comparing superbubble and blastwave SN feedback models in galaxies with stellar masses from $2 \times 10^9\ M_\odot$ to $2 \times 10^{11}\ M_\odot$ found that superbubble feedback more effectively regulated star formation and produced more disk-dominated systems (B. W. Keller et al. 2015; B. W. Keller 2022). However, with our higher resolution simulations at lower masses, the results between superbubble and blastwave seem to converge (B. Azartash-Namin et al. 2024). B. Azartash-Namin et al. (2024) directly compare superbubble and blastwave feedback methods using a subset of Marvel simulations with stellar masses ranging from $5 \times 10^3\ M_\odot$ to $10^9\ M_\odot$. Their analysis reveals that both feedback models yield similar results, showing no significant differences in either DM density slopes or the $M_*$–$M_{\rm halo}$ relation.

Since our sample focuses on lower-mass galaxies ($10^6\ M_\odot$ to $10^9\ M_\odot$), we expect the impact of SN feedback on DM shapes to be minimal (K. T. E. Chua et al. 2022). Comparing the effect of feedback on DM halo shapes, K. T. E. Chua et al. (2022) demonstrate that simulations incorporating baryonic feedback, even with vastly different implementations, consistently produce DM halo shapes that are more similar to each other than to those from simulations without baryonic feedback. This effect is most pronounced at high halo masses ($M_{\rm halo} \gtrsim 10^{12}\ M_\odot$), while at lower masses ($M_{\rm halo} \approx 10^{10}\ M_\odot$, $M_* \approx 10^8\ M_\odot$) no difference in DM shape has been found, even when comparing simulations with and without baryonic feedback. Based on this information, we anticipate that for stellar masses below $M_* \approx 10^8\ M_\odot$, the difference between superbubble and blastwave feedback would be negligible in terms of their impact on DM halo shapes, while for stellar masses up to $\approx 10^9\ M_\odot$, we anticipate only minor variations in the resultant halo shapes between these feedback models.

Furthermore, we analyzed shape measurements in one of the Marvel volumes (Storm) that has been run in otherwise identical configurations to compare between blastwave and superbubble feedback models. We find no significant differences between the models when comparing eight galaxies. Figure 10 shows the distribution of the galaxys' $S$ and $Q$, similar to Figure 3. There is considerable overlap between these distributions. A two-sample Kolmogorov–Smirnov test between the two feedback models reveals statistically indistinguishable distributions at both 1 $R_{eff}$ and 2 $R_{eff}$. At 1 $R_{eff}$, we find no significant differences in either stellar ($p = 0.66$ for $Q$, $p = 0.28$ for $S$) or DM ($p = 0.66$ for $Q$, $p = 0.98$ for $S$) shapes. Similarly, we find no significant differences at 2 $R_{eff}$ in both stellar ($p = 0.98$ for $Q$ and $S$) and DM ($p = 0.98$ for $Q$, $p = 0.66$ for $S$) shapes. However, we note that this comparison uses a smaller number than galaxies in our sample, and with a larger sample, different underlying distributions may be distinguishable.

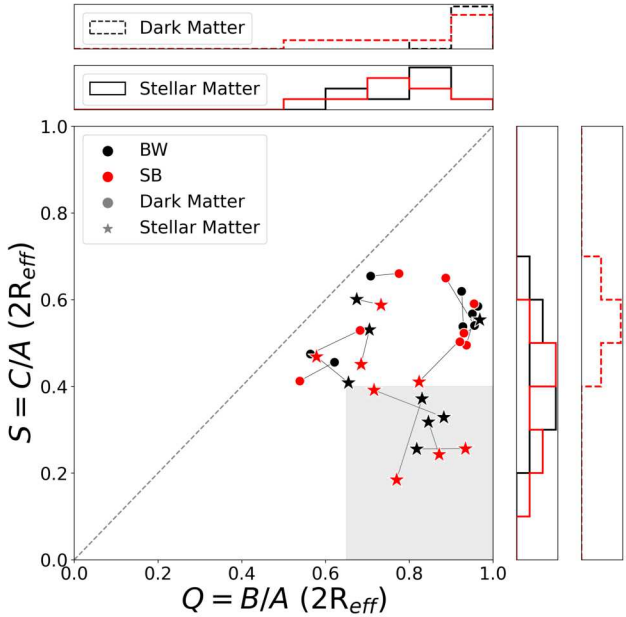

**Figure 10.** Recreation of Figure 3 comparing matched halos simulated with blastwave ("BW," black) and superbubble ("SB," red) SN feedback models in the Storm volume. Matched halos between the two feedback implementations are connected by solid lines. All other plotting conventions follow Figure 3.

These measurements at multiple radii support our decision to combine both feedback models in our analysis. Although disk formation outcomes can vary with different feedback implementations (J. Zavala et al. 2008; B. W. Keller et al. 2015, 2022; K. T. E. Chua et al. 2022), at our mass scales ($10^6 \lesssim M_*/M_\odot \lesssim 10^9$) and resolutions (see Sections 2 and 2.4), superbubble and blastwave feedback produce similar results. We are therefore confident that combining both feedback models in our sample does not significantly impact the interpretation of our results. Our data further indicate that our shape measurements are robust to specific implementations of SN feedback in dwarf galaxies.

Additionally, our simulations incorporate feedback from MBHs, although we do not expect this to significantly impact our 3D shape measurements. Previous studies in Marvel and DCJL have shown that for galaxies with stellar masses below $M_* \lesssim 10^8 M_\odot$, the MBH occupation fraction is relatively low (5%–20%; J. M. Bellovary et al. 2019). Other key properties such as the stellar mass–halo mass relation and star formation history do not show significant differences between galaxies ($10^6 \lesssim M_*/M_\odot \lesssim 10^{10}$) that host MBHs and those that do not. Finally, these MBHs typically form early in a galaxy's history ($10 < z < 20$) and accrete relatively little gas over their lifetimes, generally less than 5% of their total mass (J. M. Bellovary et al. 2019). As shown in J. M. Bellovary et al. (2019), MBHs do not contribute significantly to recent feedback processes ($z \approx 0$) in our sample of galaxies, thus minimizing their impact on our shape measurements.

### 4.3. Future Directions

Studies such as E. Kado-Fong et al. (2020) and S. Roychowdhury et al. (2013) use inference techniques to obtain the distributions of shapes of dwarf galaxies from photometry. Future work could quantify the effectiveness of these inference techniques in dwarf galaxies by simulating this inference process for a diverse sample of dwarf galaxies (S. Roychowdhury et al. 2013; E. Kado-Fong et al. 2020; W. L. Xu & L. Randall 2020). Since our simulated sample contains known shape distributions, we could directly evaluate how well inference methods recover 3D shapes, similar to W. L. Xu & L. Randall (2020), but with a larger and more diverse sample of galaxies, with a wide variety of masses, sizes, and shapes.

In conjunction with evaluating inference methods, determining how a velocity-dependent SIDM model affects the intrinsic shapes of galaxies in central regions where SIDM is expected to sphericalize the DM halo may further elucidate DM properties (R. Davé et al. 2001; P. Colín et al. 2002; A. H. G. Peter et al. 2013; V. H. Robles et al. 2017). For low-mass galaxies in particular, these differences may be detectable via alterations in the 3D shapes of stars. If stellar shape is a reliable tracer of the underlying DM shape, then we expect that a sample of nondisky low-mass galaxies ($M_* \lesssim 10^8 M_\odot$) would have the most noticeable sphericalization of stars in the inner regions of galaxies. (~1 kpc; R. Davé et al. 2001; P. Colín et al. 2002; A. H. G. Peter et al. 2013; V. H. Robles et al. 2017). If this difference is detectable, we could then apply the inference techniques to both our samples of SIDM and CDM to understand if this difference is feasible to detect solely with observations of 2D shapes.

Future planned telescopes, such as the Vera C. Rubin Observatory (Ž. Ivezić et al. 2019) and the Nancy Grace Roman Space Telescope (V. P. Bailey et al. 2023) will increase both the number of known dwarf galaxies, and also resolve them at lower masses. This process has already begun with the recently launched Euclid Space Telescope having discovered a new sample of dwarf galaxies (F. R. Marleau et al. 2025). At mass scales $M_* \approx 10^8 M_\odot$, both the Nancy Grace Roman Space Telescope and Vera C. Rubin Observatory's Legacy Survey of Space and Time (LSST) are expected to achieve high signal-to-noise ratios of 208 and 92, respectively, for lensing measurements at $r < 500$ kpc (A. Leauthaud et al. 2020). These measurements will provide a powerful tool to constrain DM halo masses and total matter distributions, including halo shapes, offering observational data to distinguish between baryonic feedback and alternate DM models (A. Leauthaud et al. 2020). Notably, the Nancy Grace Roman Space Telescope and Vera C. Rubin Observatory's LSST show the greatest potential for extending these measurements to even lower-mass scales, potentially below $M_* = 10^7 M_\odot$. The capabilities of the Nancy Grace Roman Space Telescope to make lensing measurements have been simulated and analyzed by M. A. Troxel et al. (2021), who validated the telescope's weak lensing requirements through a synthetic imaging survey. These observations will refine our data sets by extending the range of observed dwarf galaxy shapes, which could resolve tensions or uncover inconsistencies in existing DM models.

## 5. Conclusions

Using a sample of 80 high-resolution dwarf galaxies from the Marvel, DCJL, and Massive Dwarfs zoom-in simulations, we investigate the relationship between the 3D shape of SM and DM to constrain how observations of stars can be used to infer the underlying DM distribution. Our results demonstrate that in galaxies without a stellar disk (in our simulations found most commonly at $M_* \lesssim 10^{7.5}\ M_\odot$), the stellar distribution closely mirrors the underlying DM halo shape. This is evidenced by the similarity between our measurements of stellar and DM shapes (see Figures 3 and 4). We affirm that for systems without a stellar disk, the assumption holds true that the contribution of baryons to the gravitational potential is subdominant compared to the contribution of DM (e.g., C. A. Vera-Ciro et al. 2014). While we only probe masses as low as $10^6\ M_\odot$, we expect that below this mass, reduced baryon content from feedback processes (described in Section 1) would be even less influential on gravitational potentials (e.g., F. Munshi et al. 2021; M. L. M. Collins & J. I. Read 2022; D. J. Sand et al. 2024). Thus, shape measurements of stars are likely to remain good tracers of the underlying DM. In particular, we draw the following specific conclusions.

1. For galaxies without a stellar disk, we find that the shapes of SM and DM are highly correlated. This opens the potential to study properties of DM by measuring dwarf galaxy halo shapes via the stellar distributions alone.
2. Our simulated disks become thinner as stellar mass increases, over the range of masses we find them here ($10^6$–$10^{10}\ M_\odot$ in stellar mass). These trends are consistent with observations (R. Sánchez-Janssen et al. 2010; S. Roychowdhury et al. 2013; E. Kado-Fong et al. 2020).
3. Stellar triaxiality generally tracks DM halo triaxiality, with 75% of galaxies showing differences less than 0.15; this relationship is particularly strong for nondisky galaxies though weaker in disky systems which tend toward more oblate DM configurations.
4. Lower-mass, nondisky galaxies are predominantly prolate or triaxial in shape (in both SM and DM).
5. The majority of our galaxies have aligned DM and SM shapes. All galaxies with stellar disks have their SM and DM minor axes aligned, while 60% of the prolate galaxies have their SM and DM major axes aligned. Here, we define alignment to be when the axes eigenvectors are within 10° of each other.
6. When disks form, they begin to influence the shape of the DM: $Q_{DM}$ becomes rounder, $S_{DM}$ becomes thinner, and both the SM and DM becomes more oblate as measured by the triaxiality parameter.
7. The measured shape of SM and DM appears to be robust to different implementations of baryonic feedback models, showing no detectable differences using two different SN feedback models.

In conclusion, our findings suggest that for dwarf galaxies, particularly those without stellar disks or with simulated masses below $M_* \approx 10^8\ M_\odot$, the 3D shapes of stars provide a reliable proxy for the underlying DM distribution. This relationship may offer a valuable tool for inferring DM properties from stellar observations in low-mass galaxies.

## Acknowledgments

At the place George Mason University occupies, we give greetings and thanksgivings to these Potomac River life sources, to the Doeg ancestors, who Virginia annihilated in violent campaigns while ripping their lands apart with the brutal system of African American enslavement, to the recognized Virginia tribes who have lovingly stewarded these lands for millennia including the Rappahannock, Pamunkey, Upper Mattaponi, Chickahominy, Eastern Chickahominy, Nansemond, Monacan, Mattaponi, Patawomeck, and Nottaway, past, present, and future, and to the Piscataway tribes, who have lived on both sides of the river from time immemorial. Long before the University of Oklahoma was established, the land on which the University now resides was the traditional home of the "Hasinais" Caddo Nation and Kirikiris Wichita and Affiliated Tribes; more information can be found here.[8]

F.D.M. and B.K. acknowledge support from NSF grant PHY2013909. A.M.B. acknowledges support from NSF grant AST-2306340. J.D.V. was supported by the Bullard Dissertation Completion Fellowship from the University of Oklahoma. J.W. acknowledges support from an NSERC discovery grant (Natural Sciences and Engineering Research Council of Canada). F.D.M. is grateful for the hospitality of Perimeter Institute where part of this work was carried out as a Simons Emmy Noether Fellow. Research at Perimeter Institute is supported in part by the Government of Canada through the Department of Innovation, Science and Economic Development and by the Province of Ontario through the Ministry of Colleges and Universities. This research was also supported in part by the Simons Foundation through the Simons Foundation Emmy Noether Fellows Program at Perimeter Institute. This research was supported in part by grant NSF PHY-2309135 to the Kavli Institute for Theoretical Physics (KITP).

Resources supporting this work were provided by the NASA High-End Computing (HEC) Program through the NASA Advanced Supercomputing (NAS) Division at Ames Research Center. Some of the simulations were performed using resources made available by the Flatiron Institute. The Flatiron Institute is a division of the Simons Foundation. This work used Stampede2 at the Texas Advanced Computing Center (TACC) through allocation MCA94P018 from the Advanced Cyberinfrastructure Coordination Ecosystem: Services & Support (ACCESS) program, which is supported by U.S. National Science Foundation grants #2138259, #2138286, #2138307, #2137603, and #2138296.

The authors thank Ethan Nadler, Annika Peter, and Nico Garavito Camargo for useful discussions about this work. We also thank Dilys Ruan and Claire Riggs for valuable feedback on the manuscript during the drafting process.

## Data Availability

The simulation data used in this work comes from three simulation suites: "Marvel-ous Dwarfs," "DC Justice League," and "Marvelous Massive Dwarf Zooms." The "Marvel-ous

---

[8] https://www.ou.edu/cas/nas/land-acknowledgement-statement

Dwarfs" and "DC Justice League" simulation data are described in detail in F. Munshi et al. (2021), J. M. Bellovary et al. (2019), F. Munshi et al. (2019), and E. Applebaum et al. (2021).

The code used to perform our analysis will be publicly available on Github,[9] with corresponding figure data and generated $V$-band images accessible via Zenodo doi:10.5281/zenodo.15132087. The "Marvel-ous Dwarfs", "DC Justice League", and "Marvelous Massive Dwarf Zooms" simulation suites are currently proprietary. Efforts are underway to provide open access to the "Marvel-ous Dwarfs" and "DC Justice League" simulations in the future. In the interim, we welcome collaboration initiatives for researchers interested in working with these simulations. For inquiries regarding any of these simulation suites, please contact Alyson M. Brooks at abrooks@physics.rutgers.edu.

## Appendix A
## Triaxiality

Figure 11 shows a contour plot of how $Q = B/A$ and $S = C/A$ map to triaxiality. In spherical systems where $Q$ and $S$ are close to one, small changes in directions perpendicular to the contours lead to disproportional changes in triaxiality, despite reflecting largely similar shapes. Conversely, if differences in $Q$ and $S$ occur along triaxiality contours, the triaxiality values remain similar, despite having a different underlying shape.

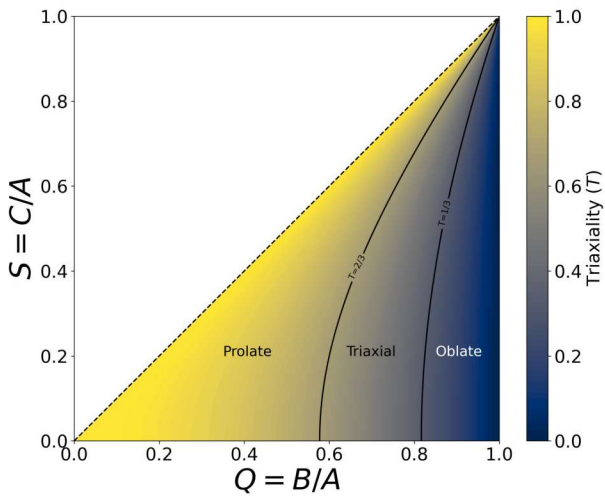

**Figure 11.** Contour plot of $T = (1 - Q^2)/(1 - S^2)$ based on values of $Q = B/A$ and $S = C/A$. The color bar shows values of triaxiality, from zero to one. The dashed line represents perfectly prolate objects at $Q = S$. While the edge of the plot at $Q = 1$ represents perfectly oblate objects. The region between the two contours at $T = 1/3$ and $2/3$ represents triaxial objects. The top right of this plot where $Q$ and $S \approx 1$, represents spherical objects, which are not well represented by a triaxiality value.

## Appendix B
## Disk Kinematics

In this appendix, we test whether our classification of disky galaxies matches the expected kinematic properties, as an additional comparison check on our definitions in Section 2.5. We find that our method of classifying galaxies based on the moment of inertia tensor is also in broad agreement with a kinematic analysis of the galaxies. We evaluate our classifications kinematically by calculating the average $J_z/J_{circ}$ ratio for star particles in each galaxy. We select star particles located between 0.5 kpc and 6 $R_{eff}$ from the galactic center for this analysis. Here, $J_z$ represents the angular momentum component perpendicular to the disk plane (after we align the disk to the $x$–$y$ plane), while $J_{circ}$ represents the angular momentum a particle would have if it followed a perfect circular orbit in the disk plane at its current radius. For a particle on a perfectly circular orbit in the disk plane, $J_z/J_{circ} \approx 1$, while values much larger than one indicate a particle is unbound. In contrast, a galaxy with nearly isotropic mass and velocity distributions instead of a disk would have an average $J_z/J_{circ} \approx 0$. Figure 12 shows the distribution of $J_z/J_{circ}$ for disky galaxies (blue) and nondisky galaxies (black), demonstrating kinematic separation between our classifications. Nondisky galaxies exhibit low $J_z/J_{circ}$ primarily within $0 \pm 0.1$ while disky galaxies show larger values of $J_z/J_{circ}$, ranging from 0.01 to 0.65. We can define a threshold at $J_z/J_{circ} = 0.05$, where below this value we expect galaxies to be nondisky, and above to be disky. With this threshold in place we measure nine (12%) galaxies that have outlier values in $J_z/J_{circ}$.

---

[9] https://github.com/Bvogel4/MorphologyMeasurements

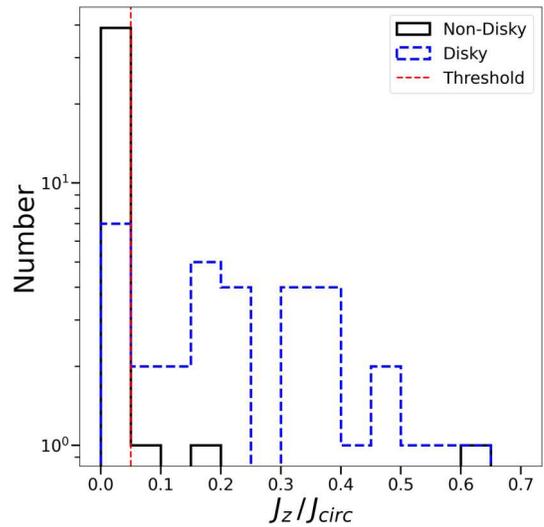

**Figure 12.** Histogram comparing the number of galaxies in each bin for the average $J_z/J_{circ}$ of stars as defined in Appendix B. The vertical axis uses a logarithmic scale to better display the wide range of galaxy counts, with bins spaced uniformly with widths of 0.05. The distribution for disky galaxies is outlined in a dashed blue line, while nondisky galaxies are outlined in a solid black line, which are classified according to the criteria in Section 2.5. A red dashed line demonstrates our separation threshold at $J_z/J_{circ} = 0.05$, where below this value we expect galaxies to be nondisky, and above to be disky. This comparison shows the separation of our disky and nondisky galaxies extends to the kinematics of our sample.

We note that there were four nondisky galaxies not shown in Figure 12 that are within $1 < J_z/J_{circ} < 3$ and do not overlap with the disky sample, one of which is a merging system. These systems may contain unbound star particles that contribute signficantly to the average $J_z/J_{circ}$.

## Appendix C
## Galaxy Environments

As shown in Figure 13, no significant difference is found in when comparing galaxy environments, i.e., isolated field dwarfs compared to satellites. The majority of satellites are categorized as nondisky (11/15). The disky satellite galaxies are found at higher dwarf masses ($M_* > 10^{8.5} M_\odot$).

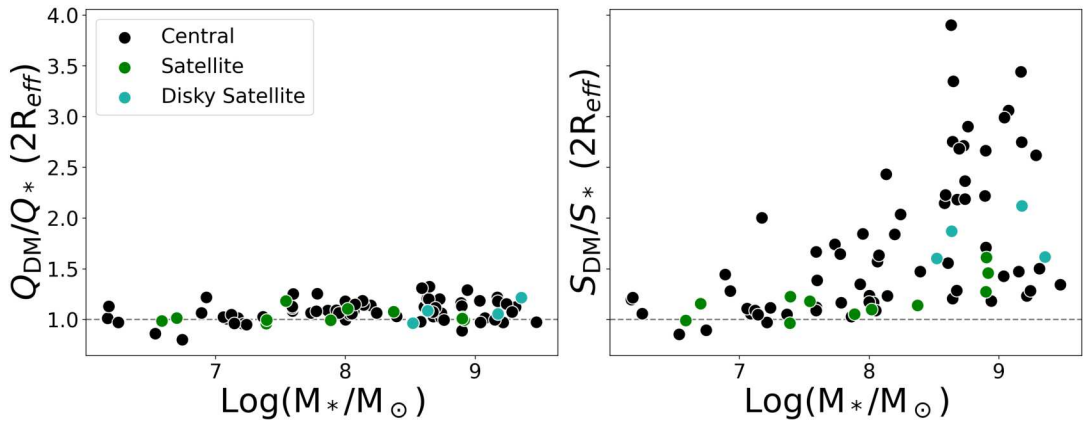

**Figure 13.** Recreation of Figure 4. Centrals are shown in black, satellites in green and cyan, with cyan representing satellites that are disky.

## Appendix D
## Shapes Measurements at Different Radii

In Figures 14 and 15 we recreate select figures at $r = 1\ R_{\rm eff}$ and make comparisons to our primary measurements taken at $r = 2\ R_{\rm eff}$. We find that galaxies with a stellar disk have less circular stellar shapes ($\bar{Q}_*$ decreases $\approx 0.1$, see Figures 3 and 14). Further, these changes in SM are reflected in triaxiality with more galaxies in the triaxial region $1/3 < T < 2/3$ (see Figure 15). We do not find notable changes in the shape of DM or stars in galaxies without a stellar disk compared to measurements at $2\ R_{\rm eff}$. Similar to our results at $2\ R_{\rm eff}$, we see considerable, but not complete overlap, with the results from E. Kado-Fong et al. (2020).

In Figure 16, we exemplify three galaxies showing their axis ratios plotted against inner radii ($0 < r < 3\ R_{\rm eff}$). These galaxies from the Marvelous Massive Dwarfs are presented in order of decreasing stellar mass $M_*$ ranging from $10^8\ M_\odot$ to $10^{9.5}\ M_\odot$. These galaxies demonstrate trends that are generally found among our sample. All of our examples show more spherical shapes in the central region ($Q = B/A$ and $S = B/A$ values closer to one). We find the thinnest disky galaxies at stellar masses closer to $10^9 M_\odot$, while lower-mass galaxies are increasingly spherical and lacking a stellar disk.

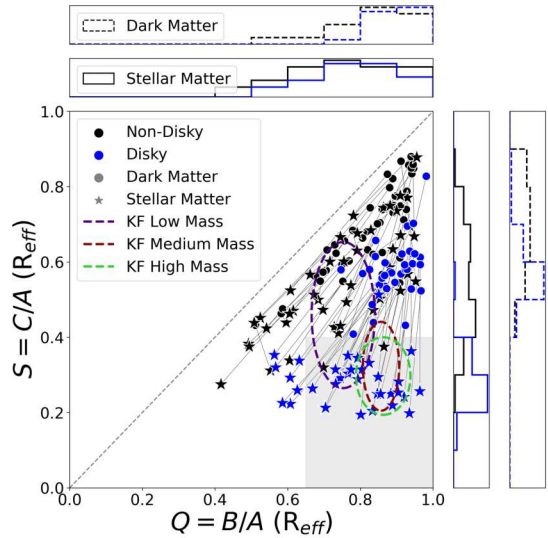

**Figure 14.** Recreation of Figure 3 but with axis ratio measurements for DM and stars at the effective radius ($R_{\rm eff}$) instead of $2\ R_{\rm eff}$.

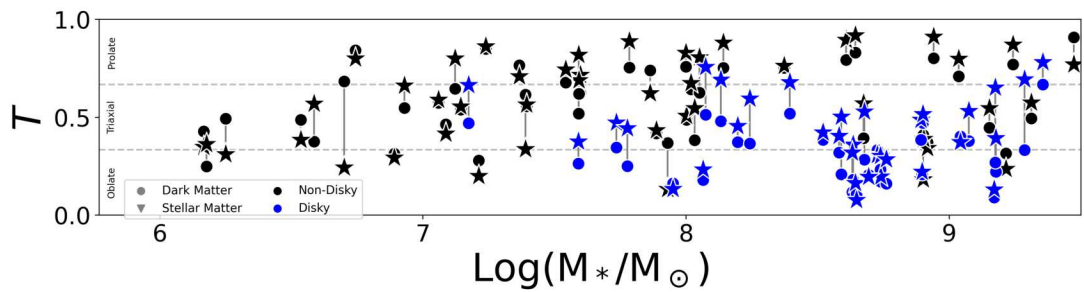

**Figure 15.** Recreation of Figure 5 but with triaxiality values for DM and stars at the effective radius ($R_{\rm eff}$) instead of $2\ R_{\rm eff}$.

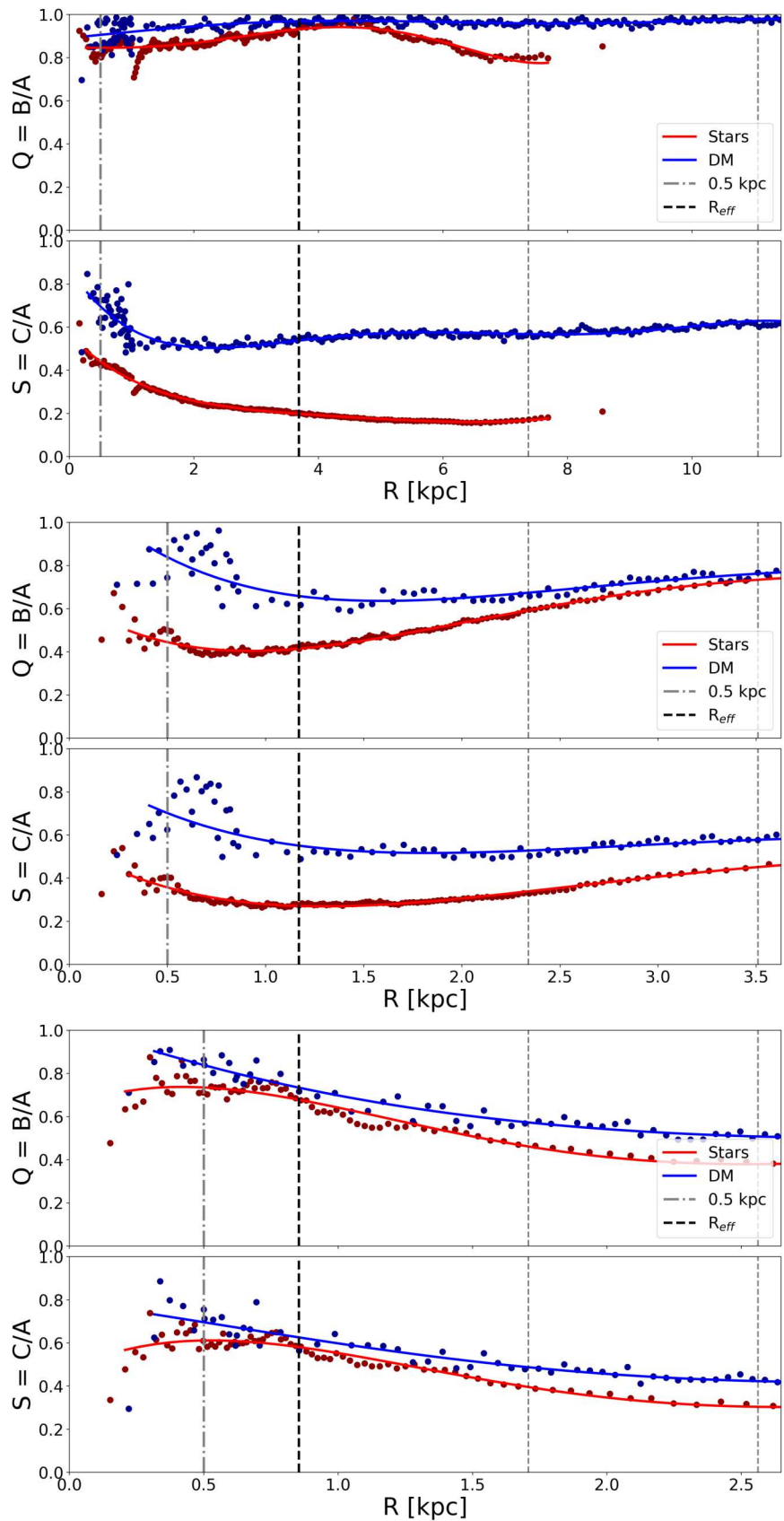

**Figure 16.** Shape profiles of three representative galaxies showing triaxiality parameters $Q = B/A$ (upper panels) and $S = C/A$ (lower panels) as a function of radius ($0 < r < 3\,R_{\mathrm{eff}}$). DM (blue) and SM (red) components are shown with raw measurements (circles) and smoothed profiles (solid lines). Vertical lines indicate key radial scales: 0.5 kpc (gray dashed–dotted), effective radius $R_{\mathrm{eff}}$ (black dashed), and multiples of $R_{\mathrm{eff}}$ at two and three (gray dashed).

## ORCID iDs

Blake Keith ● https://orcid.org/0009-0009-7745-8021
Ferah Munshi ● https://orcid.org/0000-0002-9581-0297
Alyson M. Brooks ● https://orcid.org/0000-0002-0372-3736
Jordan Van Nest ● https://orcid.org/0000-0003-3789-3722
Anna Engelhardt ● https://orcid.org/0009-0006-2183-9560
Akaxia Cruz ● https://orcid.org/0000-0001-7831-4892
Ben Keller ● https://orcid.org/0000-0002-9642-7193
Thomas Quinn ● https://orcid.org/0000-0001-5510-2803
James Wadsley ● https://orcid.org/0000-0001-8745-0263